# SEARCH FOR MUON CATALYZED ³He*d* FUSION

V.D. Fotev [a], V.A. Ganzha [a], K. A. Ivshin [a], P.V. Kravchenko, [a] P.A. Kravtsov, [a] E.M. Maev, [a]
A.V. Nadtochy, [a]  A.N. Solovyev [a],  I.N. Solovyev [a],  A.A. Vasilyev, [a]  A.A. Vorobyov, [a]
N.I. Voropaev [a], M.E. Vznuzdaev [a],  P. Kammel [b],  E.T. Muldoon, [b]  R.A. Ryan [b],  D.J. Salvat [b],
D. Prindle [b],  M. Hildebrandt [c], B. Lauss [c], C. Petitjean [c], T. Gorringe [d], R.M. Carey [e], F.E. Gray [f]

a) Petersburg Nuclear Physics Institute, Gatchina 188350, Russia
b) University of Washington, Seattle, WA 98195, USA
c) Paul Scherrer Institute, CH-5232 Villigen PSI, Switzerland d) University of Kentucky, Lexington, KY 40506, USA
e) Boston University, Boston, MA 02215, USA f) Regis University, Denver, CO 80221, USA

## ABSTRACT

This report presents the results of an experiment aimed at observation of the muon catalyzed ³He*d* fusion reaction ³He + $\mu d$ → ³He$\mu d$ → ⁴He (3.66 MeV) + *p* (14.64 MeV) + $\mu$ which might occur after a negative muon stop in a $D_2$ + ³He gas mixture. The basic element of the experimental setup is a Time Projection Chamber (TPC) which can detect the incoming muons and the products of the fusion reaction. The TPC operated with the $D_2$ + ³He (5%) gas mixture at $31K$ temperature. About $10^8$ ³He$\mu d$ molecules were produced with only 2 registered candidates for the muon catalyzed ³He*d* fusion with the expected background $N_{bg}$ = 2.2 ± 0.3 events. This gives an upper limit for the probability of the fusion decay of the ³He$\mu d$ molecule $P_F$(³He$\mu d$) ≤ 1.1·10⁻⁷ at 90% *C.L.*  Also presented are the measured formation rate of the ³He$\mu d$ molecule $\lambda_{d3He}$ = 192(3)·10⁶ s⁻¹ and the probability of the fast muon transfer from the excited to the ground state of the $\mu d$ atom $q_{1S}$ = 0.80(3).

# 1 Introduction

We report here the results of an experiment aimed at observation of the muon catalyzed ³He*d* fusion, which might occur after a negative muon stop in the $D_2$ + ³He gas mixture. The nuclear fusion reaction

$$d + {}^3He \rightarrow {}^4He\ (3.66\ MeV) + p\ (14.64\ MeV) \tag{1}$$

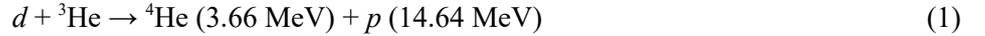

is interesting for various reasons: as a mirror reaction of the $d + t \rightarrow {}^4He + n$ fusion process and as a perspective source of thermonuclear energy. This fusion process was involved in the primordial nucleosynthesis of light elements in the early Universe. For these reasons, it is important to know the cross section of this reaction at low collision energies $E_{coll}$ < 10 keV. The phenomenon of muon catalysis of fusion reactions opens an opportunity to study this reaction at practically zero collision energy when fusion occurs in the ³He$\mu d$ muonic molecule:

$$^3He\mu d \rightarrow {}^4He\ (3.66\ MeV) + p\ (14.64\ MeV) + \mu\ . \tag{2}$$

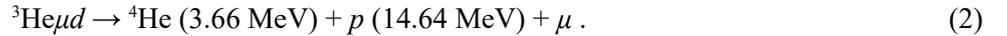

This experiment was performed in the muon beam at the Paul Scherrer institute (PSI, Switzerland).

## 1.1 Formation of the ³He$\mu d$ molecules

Formation of the ³He$\mu d$ molecules occurs in collisions of slow $\mu d$ atoms with ³He atoms:

$$\mu d + {}^3He \rightarrow [({}^3He\mu d)e]^+ + e\ . \tag{3}$$

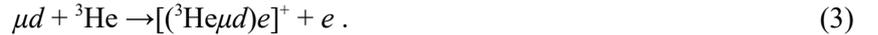

This process was predicted by Y.A. Aristov *et al*. [1] in 1981 as an intermediate step in the muon transfer from $\mu d$ to $\mu^3$He:

$$[({}^3He\mu d)e]^+ \rightarrow [(\mu^3He)e] + d + \gamma;\ \ [(\mu^3He)e] + d;\ \ \mu^3He + d + e\ . \tag{4}$$

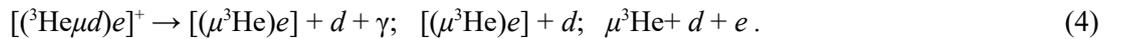

According to [1], such a scheme provides a high rate of the muon transfer from $\mu d$ to $\mu^3$He, while the direct muon transfer $\mu d \rightarrow \mu^3$He is suppressed because of the specific structure of the energy terms in the $\mu d$-³He system. This prediction was confirmed in 1993 in the experiment at the Petersburg Nuclear Physics Institute [2]. The measurements were performed with a $D_2 + {}^3He$ gas mixture at room temperature. The measured muon transfer rate $\lambda_{d3He} = (124 \pm 5) \cdot 10^6$ s$^{-1}$ proved to be in agreement with the theoretical prediction [1]. Later this muon transfer rate was measured at low temperatures in two experiments at the Paul Scherrer Institute [3,4]. The experimental results are summarized in Table 1. The theoretical predictions [1] are shown in Table 2 for comparison.

**Table 1.** Experimental results in measurements of the muon transfer rate $\lambda_{d3He}$ from $\mu d$ to $\mu{}^3He$. The density $\varphi$ is given relative to the liquid hydrogen density (LHD = $4.25 \cdot 10^{22}$ atoms cm$^{-3}$). $C_{3He}$ is the atomic concentration of ${}^3He$ in the gas mixture. The presented muon transfer rates are normalized to LHD.

| Experiment | $\lambda_{d3He} \times 10^6$ s$^{-1}$ | Experimental conditions |
|---|---|---|
| D.V.Balin et al. [2] (1993) | 124 (5) | $D_2 + {}^3He$    $T=300$ K    $\varphi = 5.5\%$    $C_{3He} = 11\%$ |
| E.M. Maev et al. [3] (1999) | 233 (16) | $HD + {}^3He$    $T=39.5$ K    $\varphi = 9.21\%$    $C_{3He} = 5.6\%$ |
| B.Gartner et al. [4] (2000) | 186(8) | $D_2 + {}^3He$    $T=30.5$ K    $\varphi = 7.0\%$    $C_{3He} = 9.13\%$ |

**Table 2.** Theoretical prediction by Y.A.Aristov et al.[1] for the rate $\lambda_{d3He}$ of the muon transfer from $\mu d$ to $\mu{}^3He$. $\varepsilon_0$ is the collision energy of the $\mu d$ atom relative to ${}^3He$. The energy $\varepsilon_0 = 0.04$ eV corresponds to $T \approx 300$ K.

| $\varepsilon_0$, eV | $4 \cdot 10^{-3}$ | $4 \cdot 10^{-2}$ | 0.1 | 1 | 10 |
|---|---|---|---|---|---|
| $\lambda_{d3He} \times 10^6$ s$^{-1}$ | 177 | 148 | 122 | 47 | 10 |

One can see that the experimental data are in agreement with the theoretical predictions, both in the absolute values and in the temperature dependence. The described scheme of the ${}^3He\mu d$ molecule formation was also supported by observation [5] of the 6.8 keV x-rays from the (${}^3He\mu d$)* decay and precision measurement of the width of this peak in agreement with the theoretical calculations [6,7].

## 1.2 Fusion process

The discovered formation process of the ${}^3He\mu d$ molecules allows to search for the muon catalyzed ${}^3He d$ fusion, similar to the muon catalyzed $dd$ and $dt$ fusion reactions. However, a serious complication arises from competition of this fusion reaction with very fast decay of the ${}^3He\mu d$ molecule through the channels shown by Eq. (4). According to the theoretical considerations [8−10], the total decay rate of the ${}^3He\mu d$ molecule is $\Lambda_{decay}({}^3He\mu d) \approx 7 \cdot 10^{11}$ s$^{-1}$. The nuclear fusion rate $\Lambda_F$ ($J$) in the ${}^3He\mu d$ molecule depends strongly on the angular momentum of the ${}^3He\mu d$ molecule. The theoretical predictions are: $\Lambda_F(J = 0) \approx 2 \cdot 10^5$ s$^{-1}$ and $\Lambda_F(J = 1) \approx 6.5 \cdot 10^2$ s$^{-1}$ [11−12]. Unfortunately, the great majority of the initially produced ${}^3He\mu d$ molecules are in the $J = 1$ state. However, as it was suggested by M. Faifman and L. Menshikov [13], the transition $({}^3He\mu d)_{J=1} \rightarrow ({}^3He\mu d)_{J=0}$ is possible in collisions of the $[({}^3He\mu d)\,e]^+$ complex with deuterium molecules via formation of a large molecular cluster $[({}^3He\mu d)\,eD_2]$ and its decay:

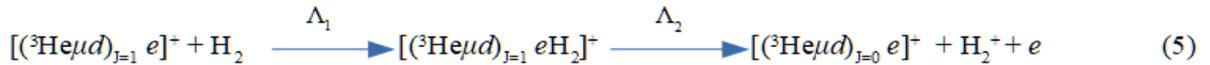

$$[({}^3He\mu d)_{J=1}\, e]^+ + H_2 \xrightarrow{\Lambda_1} [({}^3He\mu d)_{J=1}\, eH_2]^+ \xrightarrow{\Lambda_2} [({}^3He\mu d)_{J=0}\, e]^+ + H_2^+ + e \qquad (5)$$

with formation and the transfer rates of this cluster $\Lambda_1 \approx 3 \cdot 10^{13} \varphi$ s$^{-1}$ and $\Lambda_2 \approx 7 \cdot 10^{11}$ s$^{-1}$, respectively, where $\varphi$ is the H$_2$ density normalized to LHD. Here H$_2$ stands for D$_2$ or HD. Such an estimate shows that one can expect a quite efficient $({}^3He\mu d)_{J=1} \rightarrow ({}^3He\mu d)_{J=0}$ transfer and, as a consequence, a detectable ${}^3He\mu d$ fusion process. The "effective" fusion rate $\Lambda_F({}^3He\,\mu d)$ can be defined as

$$\Lambda_F({}^3He\,\mu d) = P(J=0) \cdot \Lambda_F(J=0) + P(J=1) \cdot \Lambda_F(J=1), \qquad (6)$$

where $P(J)$ is the population of the ${}^3He\mu d$ molecule state with the angular momentum $J$.

## 1.3 Previous experiments

The first experimental limit on the "effective" muon catalyzed $d^3$He fusion rate $\Lambda_F$ ($^3$He$\mu d$) ≤ 4·10$^8$ s$^{-1}$ was set at PNPI in 1990 in an experiment with the D$_2$ + $^3$He (5%) gas mixture. Next measurements were carried out in 1996 using a HD + $^3$He (5.6%) gas mixture. During a short test run in the intense muon beam at PSI, the upper limit for the $^3$He$\mu d$ fusion rate was moved down to $\Lambda_F$($^3$He$\mu d$) ≤ 1.6·10$^6$ s$^{-1}$. Later in 1997, there was a special physics run at PSI aimed at observation of the muon catalyzed $d^3$He fusion in a HD + $^3$He (5.6%) gas mixture. This experiment resulted with a new upper limit for the effective fusion rate $\Lambda_F$($^3$He$\mu d$) ≤ 6·10$^4$ s$^{-1}$ [14].

On the other hand, another collaboration at PSI has undertaken in 1998 a search of the muon catalyzed $d^3$He fusion in a D$_2$ + $^3$He (5%) gas mixture. The reanalyzed results of that experiment were published in 2006 [15]. The authors declared the first observation of this process with the measured effective fusion rates $\Lambda_F$($^3$He$\mu d$) = (4.5 + 2.6 /− 2.0) ·10$^5$ s$^{-1}$ and $\Lambda_F$($^3$He$\mu d$)= (6.9 + 3.6 /− 3.0) ·10$^5$ s$^{-1}$ at the gas mixture density φ = 5.21% and φ =16.8%, correspondingly. Such fusion rate exceeded by an order of magnitude the upper limit set in experiment [14]. This striking difference might be related with problems of taking into account the background reactions, which could simulate the searched reaction (2) in experiment [15]. The main background of this type is due to the so-called $^3$He + $d$ fusion-in-flight. It comes from collisions of the $^3$He (0.82 MeV) nuclei produced in the $d\mu d$ fusion reaction with D$_2$. This background is more important in the D$_2$ + $^3$He gas mixture than in the HD + $^3$He gas mixture used in [14]. On the other hand, the difference between the results of these two experiments might be due to possible differences in the formation and transfer rates $\Lambda_1$ and $\Lambda_2$ in the [($^3$He$\mu d$)$_{J=1}$$e$HD]$^+$ and [($^3$He$\mu d$)$_{J=1}$$e$D$_2$]$^+$ clusters.

Fortunately, the MuSun experiment, presently under way at PSI [16], gives us an excellent possibility to clarify the situation. This very high statistics experiment is using pure D$_2$ gas. This gives a possibility to control the level of the $dd$ fusion background expected in the experiment presented here.

## 2 Experimental set-up

Our experiment was performed exploiting the experimental set-up of the MuSun experiment. The main goal of MuSun is to measure the muon capture rate in deuterium. For that, the lifetime of negative muons stopped in ultra clean D$_2$ gas is measured with high precision (10$^{-5}$). That requires very high statistics of the detected muon decays. In particular, 1.054·10$^{10}$ decays of the muons stopped in the sensitive volume of the MuSun active target were registered in Run 8 of this experiment. Besides muons, the active target can detect also the products of the reactions initiated by muons, including the products of the $^3$He + $d$ fusion-in-flight :

$$^3\text{He (0.82 MeV)} + d \rightarrow {}^4\text{He (1.8−6.6 MeV)} + p \text{ (17.4−12.6 MeV)}. \qquad (7)$$

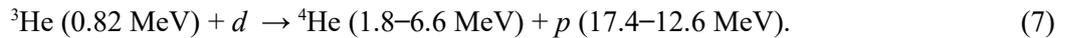

Therefore, Run 8 can serve as a high statistical background experiment for the experiment aimed at searches of the muon catalyzed $^3$He$d$ fusion in the D$_2$ + $^3$He gas mixture. Having this in mind, the decision was taken by the MuSun collaboration to perform an additional Run 9 with the active target filled with D$_2$+$^3$He (5%) gas mixture, keeping all experimental conditions identical to those in Run 8. The results of these studies are presented below.

The principal scheme of the MuSun experiment is shown in Fig.1. The incoming muons are detected first with a thin scintillator counter $\mu$SC and with a wire proportional chamber $\mu$PC. Then they pass through a 0.4 mm thick hemispheric beryllium window and stop in the sensitive volume of the time-projection chamber TPC. The TPC is the key element of the experimental set-up. It is filled with ultra-pure protium-depleted deuterium gas at $T$ = 31 $K$, $P$ = 5 bar, and it operates as an active target in the ionization grid chamber mode (without gas amplification). Its main goal is to select the muon stops within the fiducial volume of the TPC well isolated from the chamber materials.

The trajectory and the arrival time of the muon decay electrons are measured with two cylindrical wire chambers $e$PC1, $e$PC2 and with a double layer scintillator hodoscope $e$SC consisting of 32 plastic scintillators. The geometrical acceptance of the electron detector is 70%.

The ionization electrons produced in the TPC drift towards the anode plane in an electric field of 11 kV/cm with the velocity of 5 mm/$\mu s$. The total drift space (the cathode – grid distance) is 72 mm. The anode plane is subdivided into 48 pads making a pad matrix of six pads (horizontal direction $X$) by eight pads (beam direction $Z$). The size of the pads is 17.5 mm ($X$) × 15.25 mm ($Z$). About 50 % of the muons passing through the $\mu$SC are stopped within the fiducial volume above the 20 central pads at the distance of more than 1 cm from the cathode and from the grid (Fig. 2).

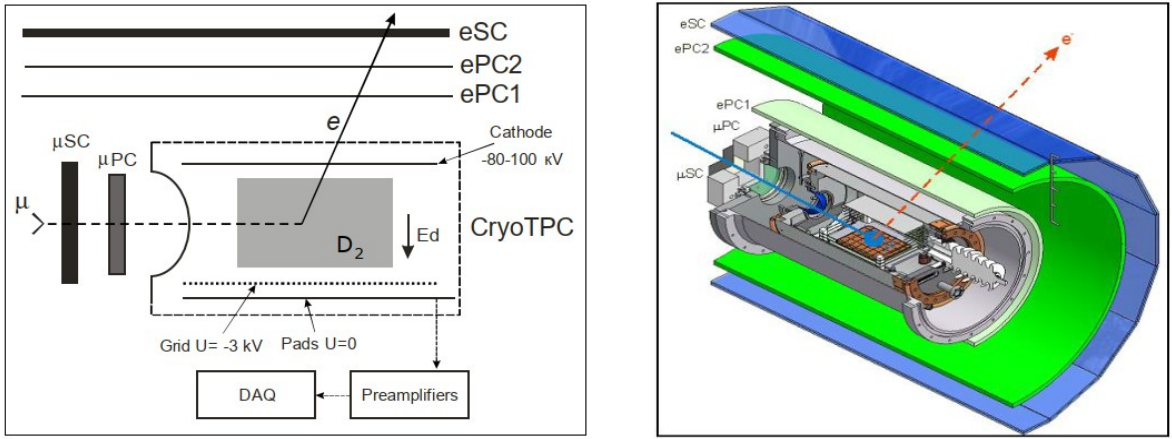

**Fig. 1.** Principal scheme of the MuSun experiment and schematic view of the MuSun set-up. The TPC is filled with ultra-pure protium-depleted deuterium gas at $T = 31$ K, $P = 5$ bar. The shadowed area shows a fiducial volume with muon stops far enough from all TPC materials.

All anode pads have independent readout channels with fast (100 MHz) ADCs allowing to measure the amplitude, the duration, the energy, and the time of appearance of the signals with the amplitude exceeding the 80 keV threshold at any pad in the time window 0−25 $\mu s$ after a muon signal is detected by the $\mu$SC counter. The same $\mu$SC signal triggers a "muon-on-request" system, which switches off the muon beam thus excluding arrivals of other muons in the registration time window. The energy resolution (noise) in each channel is around 20 keV (sigma). The TPC measures the ionization produced by the muon entering the TPC, determines its trajectory, and selects the 3D muon stop coordinate to be inside the fiducial volume isolated from all TPC materials. Also, the TPC detects products of reactions following the muon stop, including the products of the $dd$ and $^3$He$d$ fusion ($^{3,4}$He and protons).

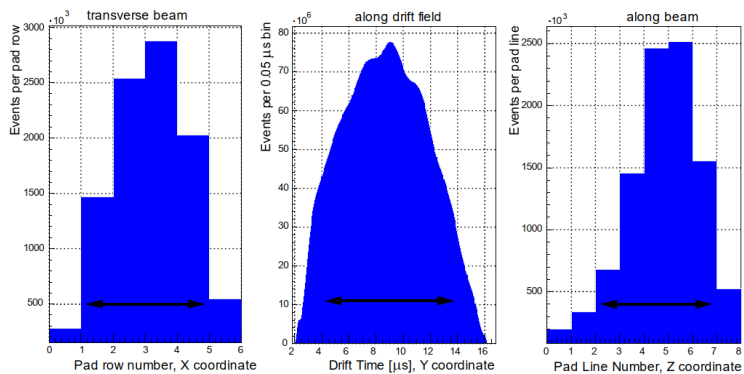

**Fig.2**. The measured muon stop distribution inside the TPC sensitive volume.
The double arrows show the fiducial volume selected in the present analysis.

# 3 Experimental data and analysis

Table 3 compares the experimental conditions of the TPC in Run 8 and Run 9. The only difference was the gas filling. All other conditions were identical. The ultra high gas purity was maintained by continuous operation of a gas purification system in both runs.

**Table 3.** Experimental conditions in Run 8 and in Run 9.

| Run | Gas filling | $T$ | $P$ | $\varphi$ | $C_d$ % | $C_{3He}$ % | Gas purity |
|---|---|---|---|---|---|---|---|
| Run 8 | $D_2$ | 31 $K$ | 5 bar | 6.5% LHD | 100 |  | $<2\cdot10^{-9}$ ($N_2$) |
| Run 9 | $D_2 + {}^3He$ | 31 $K$ | 5 bar | 6.5% LHD | 95 | 5 | $<2\cdot10^{-9}$ ($N_2$) |

## 3.1 Processes after a negative muon stop in the $D_2 + {}^3He$ gas mixture

Figure 3 shows the scheme of processes initiated by a muon stop in the $D_2 + {}^3He$ gas mixture. For the goal of this experiment, it is important to know the yield of the ${}^3He\mu d$ molecules leading to possible muon catalyzed ${}^3He d$ fusion and the yield of the ${}^3He$ (0.82 MeV) nuclei responsible for the main background reaction ${}^3He + d$ fusion-in-flight. Both yields can be calculated, as the parameters needed for such calculations are known with high accuracy. Table 4 presents the main parameters of the $d\mu d$ fusion taken from the review article [17]. Other parameters are presented in the Fig.3 caption.

**Table 4.** Parameters of the $d\mu d$ fusion at $T=31$ $K$. $\lambda_{d\mu d}(3/2)$ and $\lambda_{d\mu d}(1/2)$ are the $d\mu d$ formation rates normalized to LHD from the $\mu d$ ($F=3/2$) and $\mu d$ ($F=1/2$) states, respectively. $\lambda_{21}$ is the muon transfer rate from $\mu d$ ($F=3/2$) to $\mu d$ ($F=1/2$) normalized to LHD. $R = (\omega_{\mu 3He} + \omega_{3He})/\omega_t$. The parameters $\omega_t$, $\omega_{\mu 3He}$, and $\omega_{3He}$ are relative probabilities of the $d\mu d$ fusion channels (Fig.3)

| Parameter | $\lambda_{d\mu d}(3/2)$ | $\lambda_{d\mu d}(1/2)$ | $\lambda_{21}$ | $R(3/2)$ | $R(1/2)$ | $\omega_{\mu 3He} / (\omega_{3He} + \omega_{\mu 3He})$ |
|---|---|---|---|---|---|---|
| Value | $4.05(6)\cdot10^6 s^{-1}$ | $0.051(1)\cdot10^6 s^{-1}$ | $37.1(3)\cdot10^6 s^{-1}$ | 1.43 | 1.05 | 0.121(1) |

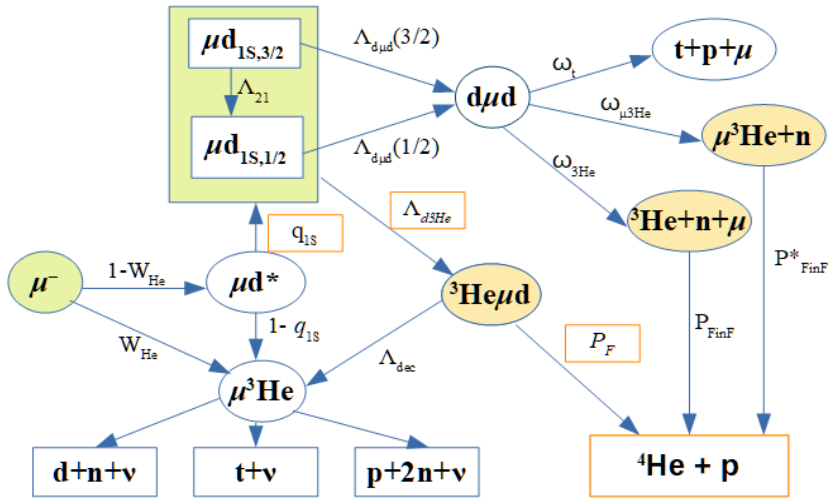

**Fig. 3.** Scheme of the processes initiated by the muon stops in the $D_2 + {}^3He$ gas mixture at $T=31K$. The parameters related to the $d\mu d$ fusion are presented in Table 4. The $d\mu d$ formation rate $\Lambda_{d\mu d} = \lambda_{d\mu d}\cdot\varphi\cdot C_d$. The spin-flip rate in the $\mu d$ atoms $\Lambda_{21} = \lambda_{21}\cdot\varphi\cdot C_d$. The parameters $\omega_t$, $\omega_{\mu 3He}$, and $\omega_{3He}$ are the relative probabilities of the $d\mu d$ fusion channels. $R = (\omega_{\mu 3He} + \omega_{3He})/\omega_t$. $P_{FinF}$ and $P^*_{FinF}$ are the ${}^3He + d$ fusion-in-flight probabilities for the ${}^3He$(0.82 MeV) and $\mu^3He$(0.80 MeV) particles, respectively. $W_{He}$ is the probability of direct muon capture by ${}^3He$. The parameter $q_{1S}$ is the probability of the muon transfer from $\mu d^*$ to $\mu d_{1S}$. $\Lambda_{d3He} = \lambda_{d3He}\cdot\varphi\cdot C_{3He}$ is the ${}^3He\mu d$ formation rate. The decay rate of the ${}^3He\mu d$ molecules $\Lambda_{dec} \approx 7\cdot10^{11}$ s$^{-1}$ [8-10]. The ${}^3He\mu d$ formation rate $\Lambda_{d3He}$, as well as the probability $q_{1S}$, are determined in the analysis of the experimental data presented here. $P_F$ is the probability of the ${}^3He\mu d$ fusion decay.

## 3.2 Energy and time distributions of the *dμd* fusion products

Figure 4 shows the energy distributions of the $^3$He and $\mu^3$He particles produced in the *dμd* fusion reaction. The largest peak in the energy spectrum of the *dd* fusion events is due to the $^3$He(0.82 MeV) particles from the *dμd* → $^3$He + n + μ fusion channel. The next peak is due to the $\mu^3$He (0.80 MeV) particles from the *dμd* → $\mu^3$He + n fusion channel. These peaks proved to be shifted from 0.82 MeV and 0.80 MeV to lower energies because of the electron-ion recombination, the effect being larger for doubly charged $^3$He$^{++}$ ions than for singly charged $\mu^3$He$^+$ ions. The difference between the peak positions in Run 8 and in Run 9 is due to difference in the effect of recombination which is by 9% lower in the D$_2$ + $^3$He (5%) gas mixture than in pure D$_2$ gas. The third peak is due to pileup of the $^3$He signals with the $^3$He or $\mu^3$He signals from the next fusion cycle.

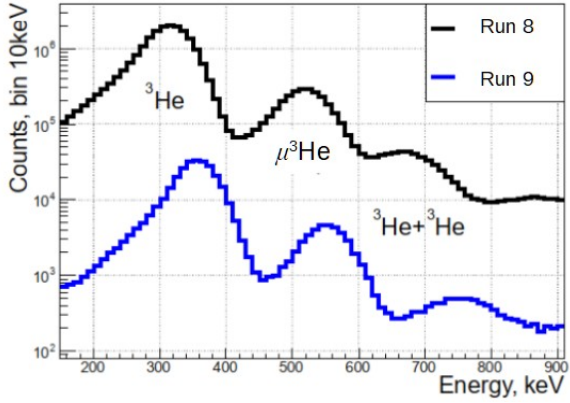

**Fig.4.** Energy distributions of the $^3$He and $\mu^3$He particles produced in the *dμd* fusion reaction. The experimental data from Run8 and Run 9 are shown with black and blue lines, respectively.

Figure 5 (left panel, blue line) presents the measured time distributions of the sum of the $^3$He and $\mu^3$He signals registered in Run 8 in the energy range from 150 keV to 620 KeV. The drop below $t = 1.5$ μs is related with overlapping of the fusion products signal with the muon stop signal. The averaged value $t^* = 1.28$ μs can be considered as the minimal time needed for separation of these signals from each other. Figure 5 (right panel, blue line) shows the time distribution of the $^3$He signals registered in Run 9 in the energy range from 200 KeV to 460 keV.

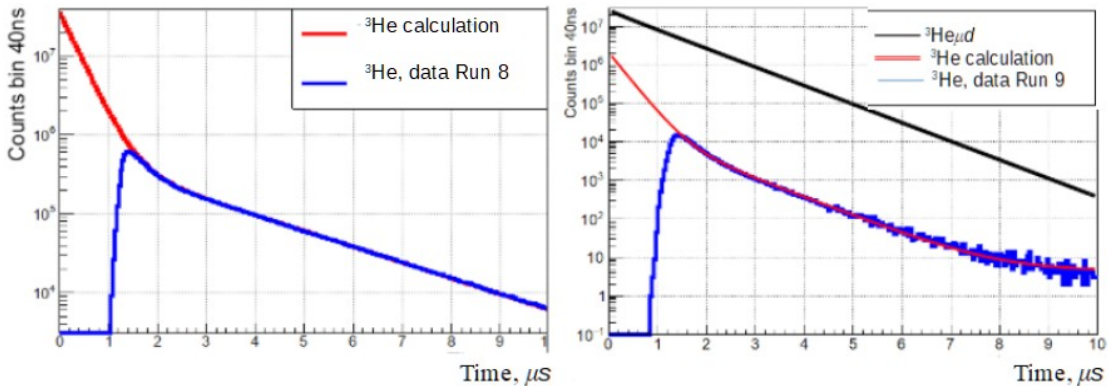

**Fig.5.** Time distributions of the $^3$He + $\mu^3$He particles from the *dμd* fusion reaction in Run 8 (blue line in the left panel) and the time distribution of the $^3$He particles in Run 9 (blue line in the right panel). The drop in the measured time distributions at $t < 1.5$ μs is due to exclusion of the fusion signals which are not enough separated from the muon stop signals. The red lines in both panels present the results of the calculations according to the scheme in Fig. 3. The black line in the right panel shows the calculated time distributions of the $^3$He$\mu$d molecules.

## 3.3 Formation rates of the *dμd* and ³He*μd* molecules and the $q_{1s}$ probability

The measured time distributions were analyzed according to the scheme presented in Fig.3. At the first step, the time distribution of the ³He + $\mu$³He events from Run8 (pure D$_2$ gas) was fitted with the calculated time distribution. For illustration, the asymptotic view (t > 2.5$\mu s$) of this distribution is presented by the following expression:

$$N_{3He+\mu 3He}(t) = N_{d\mu(1/2)} (\omega_{3He} + \omega_{\mu 3He}) \cdot K_{sel} \cdot \Lambda_{d\mu d}(1/2) \cdot \exp[-[\Lambda_0 + \Lambda_{d\mu d}(1/2)]t] . \qquad (8)$$

Here, $N_{d\mu(1/2)} = \varepsilon_{d\mu(1/2)} N_\mu$, $N_\mu = 1.054 \cdot 10^{10}$ is the number of the selected muon stops, $\varepsilon_{d\mu(1/2)} = 0.96$ is the fraction of the incoming muons reached the $d\mu(1/2)$ state, $\omega_{3He} = 45.0\%$ and $\omega_{\mu 3He} = 6.2\%$ are the probabilities of the corresponding *dμd* fusion channels, $\Lambda_0 = 0.45517 \cdot 10^6 \, s^{-1}$ is the muon decay rate, and $\Lambda_{d\mu d}(1/2) = \lambda_{d\mu d}(1/2) \cdot \varphi \cdot C_d$ is the *dμd* formation rate. $K_{sel}$ is the selection factor of the ³He + $\mu$³He fusion events. It includes rejection of the events with the energy $E < 150$ keV and $E > 620$ KeV. Also, it takes into account some losses of the events with small muon signals on the muon stop pad below some threshold value (muon stops in between the pads). The estimated value of this factor is $K_{sel} = 0.88(1)$. The rate $\lambda_{d\mu d}$ was a free parameter in these fits. To check the stability of the obtained results, the fitting procedure was repeated in various fit windows in the time interval from 2.5$\mu s$ to 10$\mu s$. The results are presented in Table 5.

**Table 5.** The *dμd* formation rate $\lambda_{d\mu d}(1/2)$ is determined from the time distribution of the ³He + $\mu$³He particles produced in the *dμd* fusion reaction. The Run 8 data (pure D$_2$ gas). The fitting procedure was repeated for various time intervals.

| Fit window, $\mu s$ | 2.5 - 10 | 2.5 - 8 | 2.5 - 6 | 2.5 - 4 | 3 - 10 | 3.5 - 10 | 4 - 10 |
|---|---|---|---|---|---|---|---|
| $\lambda_{d\mu d}(1/2)$, $10^6 \, s^{-1}$ | 0.05271(2) | 0.05268(2) | 0.05263(2) | 0.05267(2) | 0.05280(2) | 0.05280(2) | 0.05275(2) |

One can see from Table 5 that the measured *dμd* formation rate is stable within 0.2%. As an average, we take:

$$\lambda_{d\mu d}((1/2)) = 0.0527(7) \cdot 10^6 \, s^{-1} \text{ at } T = 31 \, K.$$

The error indicated here is dominated by the uncertainty in $K_{sel}$. This result agrees with the *dμd* formation rate presented in Table 4 : $\lambda_{d\mu d}((1/2)) = 0.051(1) \cdot 10^6 \, s^{-1}$. Such agreement demonstrates the validity of all parameters entering the *dμd* fusion scheme. The calculated time distribution is shown in Fig.5 ( red line, left panel). Note that these calculations reproduce the absolute yield of the ³He + $\mu$³He particles without introducing any free normalization factor.

A similar analysis of the Run 9 data (D$_2$ + ³He ) includes two more parameters. One of them is the formation rate of the ³He*μd* molecule $\Lambda_{d3He} = \lambda_{d3He} \cdot \varphi \cdot C_{3He}$. The other one is the probability $K_{fast}$ for the fast muon transfer to ³He either via the direct muon capture by ³He or in the $\mu d^*$ de-excitation process. This factor can be expressed as $K_{fast} = N_{3He\text{-fast}} / N_\mu = (1 - N_{\mu d\text{-fast}}) / N_\mu$. Here $N_{3He\text{-fast}}$ and $N_{\mu d\text{-fast}}$ are the yields of the $\mu$³He and $\mu d$ atoms after the $\mu d^*$ de-excitation process. $N_{\mu d\text{-fast}}$ is the sum of the statistically populated $\mu d$ (F=3/2) and $\mu d$(F=1/2) states (2/3 to 1/3 ratio). $\Lambda_{d3He}$ and $K_{fast}$ were free parameters in this analysis. As concerns $K_{fast}$, it was represented by $N_{\mu d\text{-fast}}$ in the fitting procedure.

The time distribution of the ³He particles calculated according to the kinetics scheme in Fig.3 was fitted to the experimental ³He time distribution in various fit windows in the time interval from 2.5$\mu s$ to 7$\mu s$. Unlike the Run 8 data, the ³He time distribution in Run 9 contains some background, though at a very low level. This background might be due to the $\mu$³He brake up reactions $\mu$³He → t+ν and $\mu$³He → d+p+ν and due to accidental noise signals. The great majority of the $\mu$³He brake up background is rejected by requirement of the muon-electron coincidence. What remains is the accidental noise signals. This background was included in the fitting program as a constant, its value being determined from the number of counts in the time interval from 8$\mu s$ to 10$\mu s$ : 4,5 events per 40 ns bin.

This analysis allowed to determine the formation rate of the ³He*μd* molecule $\lambda_{d3He}$, the fast muon to ³He transfer probability $K_{fast}$ , and the yield of the ³He*μd* molecules. The results are presented in Table 6. The calculated time distributions of the ³He particles and ³He*μd* molecules are displayed in Fig.5 (right panel) by the red and black lines, respectively.

**Table 6.** The results of analysis of the Run 9 data according to the kinetics scheme shown in Fig.3. The formation rate of the ³Heµd molecule $\lambda_{d3He}$, the probability $K_{fast}$ for the fast muon transfer to ³He, and the total number of the produced ³Heµd molecules were determined from fitting to the measured ³He time distribution in various time windows.

| Fit window, µs | 2.5 - 7 | 2.7 - 7 | 2.9 - 7 | 3.1 - 7 | 3.3 - 7 | 3.5 - 7 |
|---|---|---|---|---|---|---|
| $\lambda_{d3He} \cdot 10^6$ s⁻¹ | 194 (2) | 192(2) | 191(2) | 191(3) | 189(3) | 189(4) |
| $K_{fast}$ % | 24.5(1.8) | 26.5(2.0) | 26.8(2.3) | 27.7(2.7) | 29.0(3.0) | 29.4(3.7) |
| $\chi^2$/NDF | 1.37 | 1.37 | 1.4 | 1.4 | 1.4 | 1.36 |
| $N_{3He\mu d}$ (4π) | 0.96·10⁸ | 0.94·10⁸ | 0.93·10⁸ | 0.92·10⁸ | 0.91·10⁸ | 0.90·10⁸ |

The obtained results proved to be practically independent on the chosen fit window. We take as the final results:

$$\lambda_{d3He} = 192(3) \cdot 10^6 \text{ s}^{-1} \ ; \qquad K_{fast} = 27(3) \ \% \ ; \qquad N_{3He\mu d}(4\pi) = 0.93(3) \cdot 10^8 \text{ molecules.}$$

We can determine now the probability $q_{1S}$ of the muon transfer from µd* to µd using the following relation:

$$q_{1S} = (1 - K_{fast}) / (1 - W_{He}), \qquad (9)$$

where $W_{He}$ is the probability of direct muon capture by ³He. According to [18], $W_{He} = AC_{3He}/[1-(1-A)C_{3He}]$ with $A = 1.7$. This gives $W_{He} = 8.2\%$ for $C_{3He} = 5.0\%$. Then we have :

$$q_{1S} = 0.80(3) \ .$$

The measured formation rate of the *³Heµd* molecule proved to be in close agreement with the result of B.Gartner et al. [4] : $\lambda_{d3He} = 186(8) \cdot 10^6$ s⁻¹. The $q_{1S}$ probability was reported by M.Augsburger et al. [5] : $q_{1S} = 0.69(3)$. This value was obtained for gas density φ=7% and $C_{3He} = 9.13\%$, compared to φ = 6.5% and $C_{3He} = 5.0 \%$ in our experiment. Therefore, to be compared with our result, the value (1- $q_{1S}$) from ref.[5] should be reduced by a factor of 9.13/5.0 = 1.83. This gives $q_{1S} = 0.83(3)$. We conclude that both results are in good agreement, especially taking into account the principle difference between experimental methods used in these experiments ( measurement of absolute yield of the µd atoms in this experiment, X-ray studies in Ref. [5]).

### 3.4 Yields of the ³He(0.82 MeV) particles, ³Heµd molecules, and fusion-in-flight events

In our analysis, we normalize the yields of the ³Heµd molecules and the fusion-in-flight events to the yield of the ³He (0.82 MeV) particles (the first peak in Fig.4) in the energy range from 150 keV to 420 keV (Run 8) and from 200 keV to 460 keV (Run 9). Table 7 presents the yields of the ³He(0.82 MeV) particles and the *dµ³He* molecules registered at t ≥ 1.28 µs, together with the number of the selected muon stops.

**Table 7**. Statistics from Run 8 and Run 9. $N_\mu$ is the number of selected muon stops; $N(^3He)$ is the number of registered ³He(0.82 MeV) signals (the first peak in the energy spectra in Fig.4.); $N(^3He\mu d)$ is the number of produced ³Heµd molecules; $N_{FinF}$ is the number of expected fusion-in-flight events.

| Run | $N_\mu$ | $N(^3He)$ | $N(^3He\mu d)$ | $N_{FinF}(4\pi)$ Expected for 4π geometry |
|---|---|---|---|---|
| Run 8 | 6.3·10⁹ | 1.28·10⁷ | – | 518 ± 26 |
| Run 9 | 1.0·10⁹ | 3.34·10⁵ | 0.93·10⁸ | 14 ± 0.7 |

The measured number of the ³He (0.82 MeV) particles allows to calculate the number of the fusion-in-flight events ³He + d → ⁴He + p. The probability $F(^3He)$ to produce a FinF event by a ³He (0.82 MeV) particle stopping in the $D_2$ gas was calculated using the available ³He + d fusion cross sections in the ³He energy range below 1 MeV [19]. The calculated value is $F(^3\text{He}) = 2.7 \cdot 10^{-5}$ with 5% error. One should add to this value the weighted probabilities to produce the FinF events by the $\mu^3\text{He}$ (0.80 MeV) particles (the second peak in Fig.4 ) and by the $^3\text{He} +^3\text{He}$ pairs (the third peak in Fig.4). The weights of the corresponding peaks in Fig. 4 are:

$$w(^3\text{He}) / w(\mu^3\text{He}) / w(^3\text{He}^3\text{He}) = 0.842 / 0.125 / 0.033. \quad (10)$$

The relative fusion-in-flight probabilities for these types of events are:

$$F(^3\text{He}) / F(\mu^3\text{He}) / F(^3\text{He}^3\text{He}) = 1.0 / 2.3 / 2.0. \quad (11)$$

The probability $F(\mu^3\text{He})$ is larger than $F(^3\text{He})$ proportionally to the length of the tracks: $R(^3\text{He}) = 0.28$ mm and $R(\mu^3\text{He}) = 0.64$ mm. The total probability $F^*(^3\text{He})$ to produce a FinF event per one registered $^3\text{He}$ (0.82 MeV) signal is given by the following expression:

$$F^*(^3\text{He}) = F(^3\text{He}) \cdot [\, w(^3\text{He}) + 2.3 \cdot w(\mu^3\text{He}) + 2 \cdot w(^3\text{He}^3\text{He}) \,] / [w(^3\text{He}) \cdot \varepsilon_{3\text{He}}] = 4.04 \cdot 10^{-5}, \quad (12)$$

where $\varepsilon_{3\text{He}} = 0.95$ takes into account the losses of the $^3\text{He}$ events below the registration threshold (150 keV in Run 8 and 200 keV in Run 9). The obtained value of $F^*(^3\text{He})$ can be used to calculate the expected yield of the FinF events from the measured number of the $^3\text{He}$ (0.82 MeV) signals. The total $^3\text{He}$(0.82 MeV) yields in Run 8 and in Run 9 were found to be $N(^3\text{He}) = 1.28 \cdot 10^7$ and $N(^3\text{He}) = 3.34 \cdot 10^5$, respectively. Then, the expected yield of the $^3\text{He}d$ fusion-in-flight events in the time interval $t \geq 1.28$ $\mu s$ can be calculated as:

$$N_{\text{FinF}} = N(^3\text{He})\, F^*(^3\text{He}) = 518 \text{ events} \quad \text{in Run 8 and} \quad (13)$$

$$N_{\text{finF}} = N(^3\text{He})\, F^*(^3\text{He}) = 14 \text{ events} \quad \text{In Run 9.}$$

These numbers are presented in Table 7 with the 5% errors dominated by the error in calculations of $F(^3\text{He})$.

### 3.5 Selection of the candidates for the muon catalyzed $^3\text{He}d$ fusion events

The full data set from Run 8 and Run 9 was analyzed with the goal to identify the $^4\text{He} + p$ events produced in the muon catalyzed $^3\text{He}d$ fusion reaction. The muon stops were selected to be inside the TPC fiducial volume (Fig. 2). Also, it was required that the muon stops were accompanied by the muon decay electrons registered in the eSC electron detector in the time window 0–25 $\mu s$ after the muon stop. At the first step, the selection of the candidates for the $^4\text{He} + p$ events was done with the following criteria:

- there should be a signal at the muon stop pad P0 ($E_{P0} \geq 1.0$ MeV) separated in time from the muon signal and accompanied with two signals at a sequence of two neighbor pads P1 and P2;
- the pulses on pads P0, P1, and P2 should overlap in time to form a continuous track;
- there should be only one active P1 pad in between P0 and P2.

Figure 6 demonstrates an example of a registered candidate. The further selection of the candidates for the $^4\text{He} + p$ events was done using information on the energy deposits on pads P0, P1, and P2 taking into account the electron-ion recombination in the tracks.

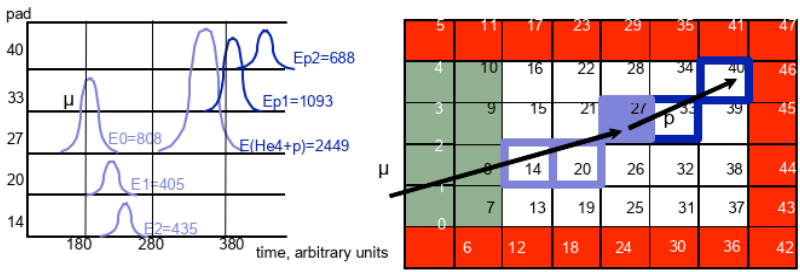

**Fig.6.** Flash ADC display of a candidate for the $^4\text{He} + p$ events. It shows a muon trajectory with the muon stop on Pad 27 followed with the signals on Pad 27, Pad 33, and Pad 40. The amplitudes of the signals shown in the figure are expressed in MeV. The pads included in the selected muon stop fiducial area are indicated with white colour.

The recombination effect reveals itself as a difference between the measured energy of the signal $E_{\text{meas}}$ and the real energy of the particle $E$: $E_{\text{meas}} = E - E_{\text{recomb}}$ (see Fig. 7b). The value of $E_{\text{recomb}}$ was determined using the measured signals from the alpha sources, $^{240}$Pu ($E_\alpha$ = 5.156 MeV) and $^{241}$Am ($E_\alpha$ = 5.480 MeV) and from the $^3$He (0.82 MeV) peak using for interpolation the following expression:

$$E_{recomb} = E (A \cdot \theta^{1/2} + B \cdot \theta), \text{ where } \Theta = Z^2 M/E. \tag{14}$$

Here $Z$ and $M$ are the charge and the mass of the ionizing particle. For Run 9, the fit parameters were found to be $A = (6.26 \pm 0.15) \cdot 10^{-3}$, $B = -(0.0095 \pm 0.0015) \cdot 10^{-3}$. In Run 8, the recombination effect is larger by a factor of 1.09.

Figure 7a shows the MC energy spectrum on pad P0 calculated for the $^3$He + $d \rightarrow$ $^4$He + $p$ fusion-in-flight events in Run 8 taking into account the recombination effect and the TPC energy resolution. Also shown is the expected energy spectrum for the muon catalyzed $^3$He$d$ fusion events in Run 9.

Figure 7c presents the energy spectrum on pad P0 of the 455 $^4$He + $p$ candidates selected in Run 8 according to the above mentioned criteria in the region $E_{P0} \geq 0.85$ MeV.

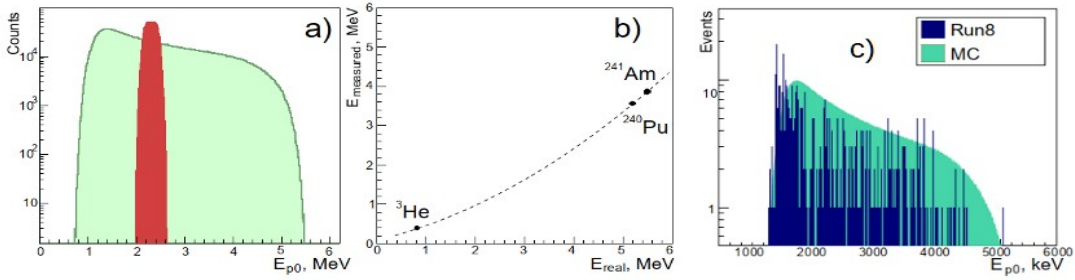

**Fig. 7.** a) MC energy spectra on pad P0 for the $^3$He + $d$ fusion-in-flight events in Run 8 (*green color*) and for the muon catalyzed $^3$He$d$ fusion events in Run 9 (*red color*). b) The measured energy versus the real energy of the $^{3,4}$He particles in Run 9. The dashed line represents the results of calculations using expression (14) with the parameters $A = 6.26 \cdot 10^{-3}$ and $B = -0.0095 \cdot 10^{-3}$ determined from the fit to the measured $^{241}$Am and $^{240}$Pu alpha peak positions and to the $^3$He (0.82 MeV) peak position. c) Energy spectrum on pad P0 of the 455 $^4$He + $p$ candidates selected in Run 8.

## 3.6 Detection efficiency and background

The next step includes the analysis of the energy spectra on pads P1 and P2. The range of the 14 MeV protons in the TPC is $R_p$ = 23 cm with $dE/dx$ = 0.35 MeV/cm. The energy deposited by a proton in the zone of pads P1 and P2 should be around 0.5 MeV. Therefore, we use the region $E_{P1} \leq 1$ MeV, $E_{P2} \leq 1$ MeV for selection of the candidates for the $^4$He + $p$ events. This resulted in 182 events in Run 8 and in 6 events in Run 9. Figure 8a displays the $E_{P1}$ x $E_{P2}$ plot of the 455 $^4$He + p candidates, with 182 events in the region $E_{P1} \leq 1$ MeV, $E_{P2} \leq 1$ MeV. One can see that, besides the $^4$He + $p$ events, there is some background with very special distribution ending sharply at $E_{P1}$ = 2.5 MeV and at $E_{P2}$ = 1.8 MeV. The nature of this background is understood. It is due to pile up of two successive $d\mu d$ fusion reactions:

$$d\mu d \rightarrow {}^3\text{H} (1.01 \text{ MeV}) + p (3.02 \text{ MeV}) + \mu, \text{ followed by} \tag{13}$$

$$d\mu d \rightarrow {}^3\text{He} (0.82 \text{ MeV}) + n (2.45 \text{ MeV}) + \mu, \text{ or vice versa.}$$

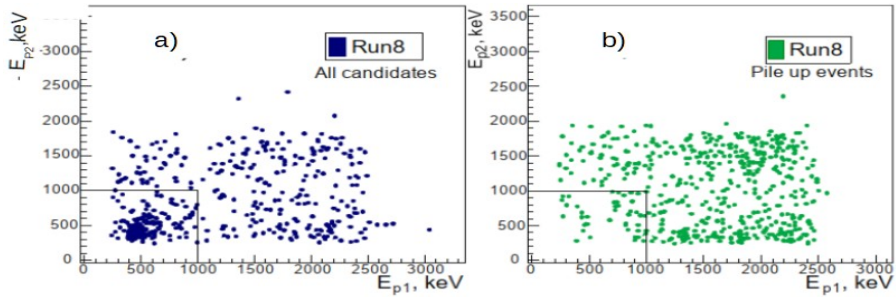

**Fig. 8.** a) Energy distribution on pads P1 and P2 of the 455 $^4$He + $p$ candidates in Run 8;
b) Energy distribution of the $dd$ fusion pileup events on pads P1 and P2 (506 events) in Run8

Such events can produce signals on P0 (due to the 1.1 MeV $^3$H and 0.82 MeV $^3$He), on P1 (due to the 3.02 MeV proton with $R_p$ = 13 mm), and on P2 (due to scattering of the 2.45 MeV neutron on deuterons). The experimental data available in Run 8 allow to reproduce directly this background by collecting the events with signals on P0 and P1 accompanied by signals on the pads which are not joining the pads P0 and P1. Figure 8b presents the $E_{P1} \times E_{P2}$ plot of such events. To separate the $^4$He + $p$ events from the $dd$ fusion pileup events, the number of events in the $E_{P1} \times E_{P2}$ plot (Fig. 8b) in the region $E_{P1} \geq$ 1 MeV, $E_{P2} \geq$ 1 MeV was normalized to the number of events in the corresponding region in Fig. 8a, and the number of the $dd$ fusion pileup events in the zone $E_{P1} \leq$ 1 MeV, $E_{P2} \leq$ 1 MeV was determined: $N_{pileup}$(R8) = 25. Then the number of the registered $d\,^3$He fusion-in-flight events was obtained: $N_{FinF}$(R8) = 182 – 25 = 157.

*Comparison of this number with the expected number of the $^3$He + d fusion-in-flight events gives the registration efficiency of the fusion-in-flight events:* $\varepsilon_F$ = 157/ 518 = (30 ± 3)%. This value is valid also for the registration efficiency of the muon catalyzed $d\,^3$He fusion. The quoted error is determined by the error in the number of the detected fusion-in-flight events and by the error in the calculated probability to produce a fusion-in-flight event by the 0.82 MeV $^3$He particle in the $D_2$ gas.

The two types of events considered above constitute the main background in Run 9 aimed at observation of the muon catalyzed fusion reaction $^3$He$d \to {}^4$He + $p$. Based on the results obtained in Run 8, we can calculate the expected background in Run 9 using the following expressions :

$$N_{FinF}(R9) = N_{FinF}(R8) \cdot N(^3He)_{R9} / N(^3He)_{R8} \cdot C_d(R9)/C_d(R8), \qquad (14)$$

$$N_{pileup}(R9) = N_{pileup}(R8) \cdot N(^3He)_{R9} / N(^3He)_{R8} \cdot P_{pileup}(R9)/P_{pileup}(R8), \qquad (15)$$

$$N_{bgr}(R9) = N_{FinF}(R9) + N_{pileup}(R9), \qquad (16)$$

where the ratio of the registered $^3$He signals $N(^3He)_{R9} / N(^3He)_{R8}$ = 0.026 ; the ratio of the $D_2$ densities $C_d(R9)/C_d(R8)$ = 0.95; and the ratio of the $dd$ fusion pileup probabilities $P_{pileup}(R9)/P_{pileup}(R8)$ = 0.61. The background predictions for Run 9 calculated in this way are as follows: $N_{FinF}$(R9) = 3.87 ± 0.3, $N_{pileup}$(R9) = 0.39 ± 0.08, $N_{bgr}$(R9) = 4.3 ± 0.4. The quoted error in $N_{bgr}$(R9) is determined mostly by the statistical error in $N_{FinF}$(R8).

We can further reduce $N_{bgr}$(R9) by cutting the low energy part in the energy spectrum on pad P0 presented in Fig. 8a. The expected position of the signals from the muon catalyzed $^3$He$d \to {}^4$He + $p$ reaction is above $E_{P0}$ = 2.4 MeV (Fig. 7a ). Therefore, we can set the low energy cut at the energy up to $E_{P0}$ = 2.0 MeV without noticeable decrease in the registration efficiency of the muon catalyzed $^3$He$d \to {}^4$He + $p$ reaction. Table 8 presents the background predicted for Run 9 for various $E_{P0}$ cuts.

Another source of background in Run 9 might be the break up reaction $\mu ^3$He$\to d + n$, $p + 2n$. However, the energy deposit on pad P0 being rather small in such events, they could simulate the muon catalyzed $^3$He$d \to {}^4$He + $p$ fusion events only when piling up with the $dd \to {}^3$H + $p$ events. The calculated probability of such process is $0.5 \times 10^{-7}$ per muon stop. In addition, it is suppressed by three orders of magnitude to a negligible level by requiring detection of the muon decay electron with the $e$PC/$e$SC detectors. Similarly, the muon capture on gas impurities ($N_2$) could be disregarded, especially taking into account very high purity ($10^{-9}$) of the $D_2$ gas in this experiment.

*Finally, two candidates for the muon catalyzed $^3$He$d$ fusion were registered with the predicted background of 2.2 ± 0.3 events.*

**Table 8.** Total number of selected $^4$He + $p$ candidates $N_{tot}$ in Run 8 and in Run 9, the number of fusion-in-flight events $N_{FinF}$, and the number of $dd$ fusion pile up events $N_{pileup}$ registered in Run 8 and extrapolated to Run 9 for various cuts on the energy deposited on pad P0.

| $E0$ cut MeV | Run 8 | | | Run 9 | Run 9 background determined from Run 8 | | |
|---|---|---|---|---|---|---|---|
| | $N_{tot}$ | $N_{FinF}$ | $N_{pileup}$ | $N_{tot}$ | $N_{FinF}$ | $N_{pileup}$ | $N_{bgr} = N_{FinF} + N_{pileup}$ |
| 1 | 182 | 157 | 25 | 6 | 3.87 ± 0.31 | 0.39 ± 0.08 | 4.3 ± 0.4 |
| 1.6 | 117 | 93 | 24 | 3 | 2.30 ± 0.23 | 0.37 ± 0.08 | 2.7 ± 0.3 |
| 2 | 99 | 77 | 22 | 2 | 1.90 ± 0.22 | 0.34 ± 0.07 | 2.2 ± 0.3 |

## 3.7 Upper limit for fusion decay probability of the $d\mu^3$He molecules

Based on this observation, an upper confidence limit for the number of the muon catalyzed $^3$He$d$ fusion events was calculated by the method described in Refs. [20,21] which takes into account the measured background uncertainty: $N_F \leq 3.1$ events at the 90% confidence level. This determines an upper limit for the probability for the fusion decay of the $^3$He$\mu d$ molecule in the $D_2 + {}^3$He(5%) gas mixture at 6.5% LHD density:

$$P_F({}^3\text{He}\mu d \rightarrow {}^4\text{He} + p + \mu) = N_F / [N({}^3\text{He}\mu d) \cdot \varepsilon_F], \quad (17)$$

where $N_F \leq 3.1$ events is the upper limit at 90% C.L. for the number of detected muon catalyzed $^3$He$d$ fusion events, $N({}^3\text{He}\mu d) = 0.93 \cdot 10^8$ is the number of produced $^3$He$\mu d$ molecules, and $\varepsilon_F = 0.30$ is the detection efficiency for the $^3$He$d$ fusion events. This gives:

$$P_F({}^3\text{He}\mu d \rightarrow {}^4\text{He} + p + \mu) \leq 1.1 \cdot 10^{-7}.$$

*One should stress here that the obtained result is model independent.* It relies only on the experimentally measured parameters including the fusion-in-flight background and the detection efficiency.

Using the measured $^3$He$\mu d$ fusion decay probability and involving the theoretical value for the total $^3$He$\mu d$ total decay rate, $\Lambda_{dec} = 7 \cdot 10^{11}$ s$^{-1}$, one can deduce the "effective" $^3$He$\mu d$ fusion decay rate:

$$\Lambda_F = \Lambda_{dec} \cdot P_F \leq 7.7 \cdot 10^4 \text{ s}^{-1} \text{ at 90% C.L.} \quad (18)$$

## 4 Conclusion

This experiment (MuSun-Run9) was aimed at the search of the muon catalyzed $^3$He$d$ fusion in the $D_2+{}^3$He (5%) gas mixture. An important advantage of these measurements was a possibility to determine the level of the background and the registration efficiency using data from the high statistical MuSun-Run8 experiment performed with pure $D_2$ gas in the same experimental conditions. This allowed to determine in a model independent way the upper limit for the probability $P_F$ of the fusion decay of the $^3$He$d\mu$ molecule in the $D_2 + {}^3$He(5%) gas mixture at 6.5% LHD density at 31K temperature:

$$P_F({}^3\text{He}\mu d \rightarrow {}^4\text{He} + p + \mu) \leq 1.1 \cdot 10^{-7}.$$

Using this value of $P_F$ and involving the theoretical value for the total $^3$He$d\mu$ total decay rate, $\Lambda_{dec} = 7 \cdot 10^{11}$ s$^{-1}$, we deduce an upper limit for the "effective" $^3$He$d\mu$ fusion decay rate $\Lambda_F$:

$$\Lambda_F \leq 7.7 \cdot 10^4 \text{ s}^{-1} \text{ at 90% C.L.}$$

This limit for $\Lambda_F$ obtained with the $D_2 + {}^3$He(5%) gas mixture is close to the result of the experiment [14] performed with the HD + $^3$He (5.6%) gas mixture: $\Lambda_F \leq 6.0 \cdot 10^4$ s$^{-1}$. On the other hand, it disagrees strongly with the rate $\Lambda_F \approx 6 \cdot 10^5$ s$^{-1}$ reported in [15] in the $D_2 + {}^3$He(5%) gas mixture, and thus rules out the statement made in [15] on observation of the muon catalyzed $^3$He$d$ fusion. This excludes also as a possible explanation of the difference in $\Lambda_F$, observed in [14] and [15], an assumption of a large difference in the formation rates (according to Eq.5) of the clusters $[({}^3\text{He}\mu d)eD_2]^+$ and $[({}^3\text{He}\mu d)eHD]^+$.

We present here also two complimentary results obtained in this experiment:
- the formation rate of the $^3$He$\mu d$ molecule: $\lambda_{d3He} = 192(3) \cdot 10^6$ s$^{-1}$;
- the probability of the muon transfer from $\mu d^*$ to $\mu d$: $q_{1S} = 0.80(3)$.

These results proved to be in good agreement with measurements reported in [4] and [5].

Based on the theoretical predictions [10–13], one could expect the rate $\Lambda_F \approx 2.5 \cdot 10^4$ s$^{-1}$, which would correspond to observation of only 1.0 event in the MuSun-Run9 experiment. Note, however, that this experiment was performed as a supplement to MuSun-Run8 with only one week running time. In a dedicated experiment with the same setup, one could increase the sensitivity for detection of the muon catalyzed $^3$He$d$ fusion events by an order of magnitude with reduced and well controlled background. This would allow to reach the theoretically predicted level.

**Acknowlegments**
The authors express their gratitude to M.P.Faifman and L.N.Bogdanova for fruitful discussions of the theoretical aspects of the muon catalyzed $^3$He$d$ fusion.
This work was supported by the Ministry of Science and Education of the Russian Federation and by the US Department of Energy Office of Science, Office of Nuclear Physics under Award No. DE-FG02-97ER41020.